\documentclass[%
 aip,
 jmp,%
 amsmath,amssymb,
nofootinbib,
preprint,%
]{revtex4-1}

\usepackage{graphicx}
\usepackage{dcolumn}
\usepackage{bm}

\def\VEC#1{\mbox{\boldmath $#1$}}

\begin{document}

\preprint{AIP/123-QED}

\title[Group velocity and causality in standard resistive RMHD]
{Group velocity and causality in standard relativistic resistive MHD} 

\author{Shinji Koide}
\affiliation{Department of Physics, Kumamoto University, 
2-39-1, Kurokami, Kumamoto, 860-8555, JAPAN}
\email{koidesin@sci.kumamoto-u.ac.jp} 

\author{Ryogo Morino}
\affiliation{RKK Computer Service Company,
1-5-11, Kuhonji, Kumamoto, 862-0976, JAPAN}
\email{morino@rkkcs.co.jp} 

\date{\today}

\begin{abstract}
Group velocity of electromagnetic waves in plasmas 
derived by standard relativistic resistive MHD (resistive RMHD)
equations is superluminal.
If we assume that the group velocity represents the propagation velocity
of a signal, we have to worry about the causality problem. That is, some
acausal phenomena may be induced, such that information transportation 
to the absolute past and spontaneous decrease in the entropy. 
Here, we tried to find the acausal phenomena using standard
resistive RMHD numerical simulations in the suggested situation 
of the acausal phenomena.  The calculation results showed 
that even in such situations no acausal effect happens.
The numerical result with respect to the velocity limit of the
information transportation is consistent with a linear theory
of wave train propagation.
Our results assure that we can use these equations without 
problems of acausal phenomena.
\end{abstract}

\maketitle


\section{Introduction} \label{sec1}

Recently, relativistic ideal MHD (ideal RMHD) equations have been used more 
frequently to perform numerical simulations of plasmas around black holes 
\citep{koide06,mckinney06}.
The ideal RMHD numerical simulations showed that around a rapidly rotating
black hole, anti-parallel magnetic field configuration, 
where the magnetic reconnection
will be caused spontaneously, is formed naturally. 
Furthermore, they showed that much
electromagnetic energy is transported into the region between the accretion disk
and the black hole ergosphere, stored in the region, and then
will be released due to the magnetic instability or magnetic reconnection.
To perform calculations of the magnetic reconnection around
black holes, we have to use the ``resistive" RMHD equations, including
resistivity, but no one succeeded in such a calculation yet. 
This is because we worry
that ``standard" equations of the resistive RMHD with the ``standard"
relativistic Ohm's law may have the causality problem, and
``generalized" RMHD equations \citep{koide10}, where the group velocity
of the electromagnetic wave in the plasma with the plasma parameter
larger than unity is smaller than the light speed $c$, are very complex
to perform numerical calculations.
That is, with respect to the standard equations, we had a concern that 
the group velocity $v_{\rm g}$
of the electromagnetic wave becomes larger than the light speed $c$ and it 
destroys causality if we assume the group velocity means the propagation 
speed of transportation of information or energy \citep{jackson62}. 
For example, it is suggested that
in a coordinates frame where the plasma flows with the velocity larger than
the critical velocity $c^2/v_{\rm g}$, 
the amplitude of the wave packet increases
spontaneously \citep{koide08}. This kind of instability contradicts to 
causality and the second law of the thermodynamics.
On the other hand, using a linear theory of electromagnetic wave
propagation, we can prove that no signal propagates with a velocity
greater than the light speed in a plasma and such acausal phenomenon
never happens as discussed by \citet{jackson62}
(see Appendix \ref{appendvwf}).

To solve confusion with respect to causality of the standard
resistive RMHD equations, we calculated expected situations of 
such acausal phenomena (instability) using the standard resistive RMHD 
equations in a flat space-time. 
In conclusion, the numerical calculations
showed that such acausal phenomena never happen. That is, in such cases
that acausal phenomena are expected, the edge of the wave packet becomes
dominant compared to the bulk part of the wave packet and the edge of the wave
propagates within the light speed. This means that superluminal group
velocity does not mean breakdown of causality as discussed before 
\citep{koide08}. 
That is, the standard resistive RMHD equations
permit the superluminal propagation of the infinite wave packet but not
superluminal information propagation, thus do not break causality.
We can use the standard resistive RMHD equations within causality.
It is noted that in a real plasma, group velocity of the electromagnetic
wave is equal to or less than the light speed due to the inertia effect 
and momentum
of electrons or electric current, which are neglected in the standard
resistive RMHD. To treat these effects, we have to use
``generalized" RMHD equations \citep{koide09,koide10}.

In this paper, we report numerical calculations of electromagnetic
wave packets using standard resistive RMHD equations.
In Section \ref{sec2}, we introduce the standard resistive RMHD equations and
explain the causality problem suggested before.
In Section \ref{sec3}, we show a numerical simulation of the electromagnetic
wave in uniform, rest, resistive plasma, and in Section \ref{sec4}, we
try to calculate the propagation of the electromagnetic wave packet
in the situation where acausal phenomena are expected.
In Section \ref{appendvwf}, with respect to one of the causality problems,
we clarify we should use the head velocity as the velocity limit of the
electromagnetic wave in plasma, not the group velocity based on
the linear theory of propagation of electromagnetic plane wave.
In Section \ref{secdiscuss}, 
we discuss validity of the standard resistive RMHD equations
for astrophysical relativistic plasmas.
In Section \ref{sec5}, we summarize the results of the numerical calculations
and present some remarks.

\section{Standard resistive RMHD and suggested causality problems \label{sec2}}

The standard resistive RMHD equations with 4-form in Minkowski spacetime are as follows:
\begin{eqnarray}
\partial_\nu (\rho U^\nu) &=& 0 , \label{onefluidnum} \\
\partial_\nu T^{\mu\nu} &=& 0 , \label{onefluidmom}  \\
\partial_\nu \hspace{0.3em} ^*F^{\mu\nu} &=& 0 , \label{4formfar} \\ 
\partial_\nu F^{\mu\nu} &=& J^\mu , \label{4formamp} \\
U_\nu F^{\mu\nu} &=& =\eta (J^\mu - \rho_{\rm e}' U^\mu), \label{4formohm}
\end{eqnarray}
where $\partial_\nu$, $U^\nu$, and $J^\nu$ are the 4-derivative in Minkowski
spacetime, 4-velocity,
and 4-current density, $F^{\mu\nu}$ is the field strength tensor, 
$^*F^{\mu\nu}$ is the dual tensor of $F^{\mu\nu}$, and
 $T^{\mu\nu}$ is the energy-momentum tensor,
\begin{eqnarray} 
T^{\mu\nu} &\equiv& p g^{\mu\nu} + h U^\mu U^\nu 
+ {F^\mu}_\sigma F^{\nu\sigma} - \frac{1}{4} g^{\mu\nu} F^{\kappa\lambda} F_{\kappa\lambda} ,
\end{eqnarray}
(see \citet{jackson62,weinberg72}).
The scalar variables, density $\rho$, pressure $p$, enthalpy density $h$,
charge density $\rho_{\rm e}' = -U_\nu J^\nu$
are observed by the rest frame of the plasma.
Equation (\ref{4formohm}) presents the standard relativistic Ohm's law.
Through this paper, we use units so that the light speed is unity, $c=1$.

\begin{figure}
\includegraphics{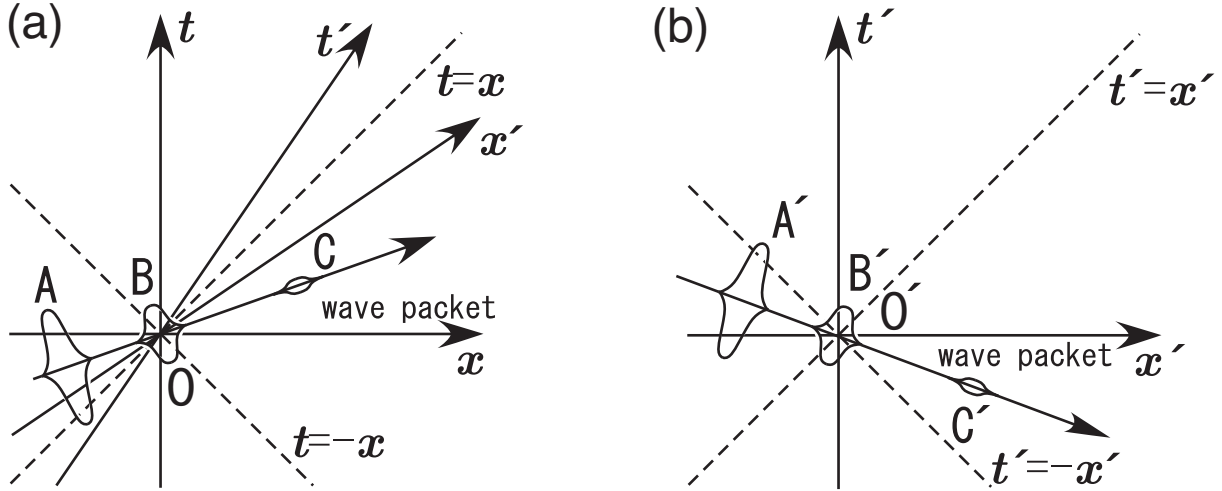}
\caption{Propagation of superluminal electromagnetic wave packet in uniform,
unmagnetized plasma. (a) In the rest plasma coordinates frame $O-xyz$, the wave 
packet propagates and damps as shown by the profiles at the points A, B, and C.
(b) In the coordinates frame with the relative velocity to the rest frame,
$v_0 > 1/v_{\rm g}$, ($O'-x'y'z'$)
A, B, and C in the panel (a)  
correspond to the points A', B', and C', respectively.
In this frame $O'-x'y'z'$, the wave packet seems to propagate 
opposite direction of the wave in the plasma rest frame and grow up.
\label{fig1}}
\end{figure}

The 3+1 formalism of the standard resistive RMHD equations is written as,
\begin{equation}
\frac{\partial \gamma \rho}{\partial t} = -  \sum _i
\frac{\partial}{\partial x^i} \gamma \rho v^i ,
\label{cmma}
\end{equation}
\begin{equation}
\frac{\partial P^i}{\partial t} = - \sum _j
\frac{\partial}{\partial x^j} T^{ij} 
\end{equation}
\begin{equation}
\frac{\partial \epsilon}{\partial t} =  -  \sum _i
\frac{\partial}{\partial x^i} \left (
P^{i} -\gamma \rho v^i \right ) 
\end{equation}
\begin{equation}
\frac{\partial B_i}{\partial t} = - \sum _{j,k}
\epsilon ^{ijk} \frac{\partial}{\partial x^j}
E_k ,
\label{cmfa}
\end{equation}
\begin{equation}
J^i + \frac{\partial E_i}{\partial t} =
\sum _{j,k}  \epsilon ^{ijk}
\frac{\partial}{\partial x^j} B_k  ,
\label{cmam}
\end{equation}
\begin{equation}
 \sum _{i}  \frac{\partial}{\partial x^i}
 B_i = 0    ,
\end{equation}
\begin{equation}
\rho _{\rm e} = \sum _{i}  
\frac{\partial}{\partial x^i}  E_i ,
\end{equation}
\begin{eqnarray}
E^i + \sum_{j,k} \epsilon_{ijk} v^j B_k = \frac{\eta}{\gamma} 
(J^i -\rho_{\rm e}' U^i)  ,
\label{cmohm}
\end{eqnarray}
where $P^i = T^{0i}$ is the 3-momentum density, $\epsilon = T^{00}$ is the energy density, 
$B^i = \frac{1}{2} \sum_{jk} \epsilon^{ijk} F_{jk}$ is the magnetic field,
$E^i = F^{0i}$ is the electric field, $\gamma =U^0$ is the Lorentz factor, 
$v^i = U^i/\gamma$ is the 3-velocity, and $\rho_{\rm e} = J^0$ is the charge density.
Using linearized equations of Eqs. (\ref{cmfa}) -- (\ref{cmohm}), we have 
the dispersion relation of the electromagnetic wave propagating
in a uniform, rest plasma:
\begin{equation}
\omega^2 - \frac{i}{\eta} \omega - k^2 =0,
\label{dispersion_relation}
\end{equation}
where $\omega$ is the angular frequency and $k$ is the wave
number of the electromagnetic wave. Here, we assumed the
resistivity is constant over the whole plasma.
Then, the group velocity of the electromagnetic wave propagating
in the uniform plasma is given by
\begin{equation}
v_{\rm g} = \frac{2\eta k}{\sqrt{(2\eta k)^2-1}}  > 1. 
\end{equation}
We found that the group velocity is always larger than the light
speed.
If we assume that the group velocity is the propagation velocity
of information, we can send an information faster
than the light speed. That is, this information propagates
outside of the causality corn $t=\pm x$ (Fig. \ref{fig1}(a)).
Furthermore, when we consider a new inertia 
coordinates frame $O'-x'y'z'$, which moves with the velocity $v_0 > 1/v_{\rm g}$
observed by the plasma rest frame $O-xyz$, the signal seems to
propagate toward the past ($t'<0$, but not ``absolute" past) (Fig. \ref{fig1}(b)).
If we combine this type of information transport in the two streaming plasmas
with the relativistic velocity $v_0 > 1/v_{\rm g}$, we may be able to
send the information to the absolute past.
This is strange because if it is true, we can 
know the today's information (for example, about stock rate) yesterday. 
This is one type of
causality problems discussed before.  The other type of causality problem is 
about  spontaneous decrease in the entropy.
Fig. \ref{fig1}(a) shows the electromagnetic wave packet propagation
with damping. 
The damping rate is approximately given by $\Im \omega = 1/(2 \eta)$
from Eq. (\ref{dispersion_relation}), when $k > 1/(2 \eta)$.
Here, let me note that the electromagnetic wave and then wave packet 
do not propagate
in the ideal MHD case ($\eta = 0$) because the damping rate is infinite 
($\Im \omega \longrightarrow \infty$).
This damping is due to the convergence of 
electromagnetic energy to thermal energy of plasma. 
During the damping of the wave packet, the entropy of the system increases.
On the other hand, when we see the same wave in the new coordinates frame 
$O'-x'y'z'$ (Fig. \ref{fig1}(b)),
we find the electromagnetic wave propagates toward the left direction
and grows up.
The increased electromagnetic energy of the grown wave packet 
comes from the thermal energy of the plasma. 
This process shows the spontaneous decrease in the entropy.
This is also a strange phenomenon. If it is true, using this effect, we can
solve the energy problem of the world.
As shown above, the causality problem is categorized into the following two types:
\begin{enumerate}
\item Transport of information toward the absolute past.
\item Spontaneous decrease in the entropy.
\end{enumerate}
In this paper, we investigate these problems 
of the standard resistive RMHD using numerical simulations.
Basically, we use the method developed by \citet{watanabe06}

\section{Information transportation \label{sec3}}

\subsection{Propagation of a simple wave packet}

\begin{figure}
\includegraphics{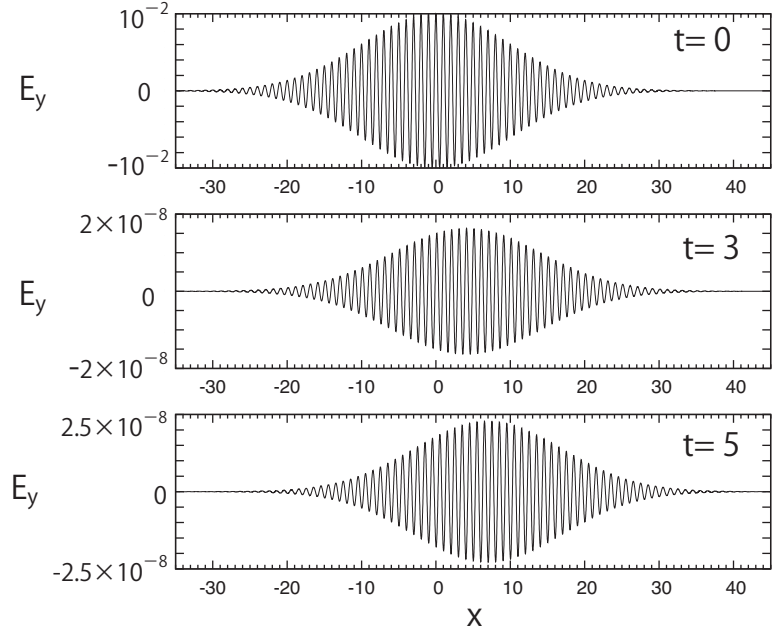}
\caption{
Simulation of propagation of a simple wave packet with a narrow 
Fourier spectrum.
\label{fig2}}
\end{figure}

We consider a simple wave packet propagation with
group velocity $v_{\rm g}$ in a rest, uniform plasma.
Here, we take the parameters of the wave packet as,
\[
\sigma = 50, \verb!   !
k_0= 2 \pi, \verb!   ! \eta = \frac{1}{2^{3/2} \pi}, \verb!   ! 
\omega_0=\frac{\pi}{\sqrt{2}} ( \pm 1 - i ) = \pm \pi e^{\mp i \pi/2},
\]
where $\sigma$ is the characteristic length of the wave packet,
$k_0$ is the typical wave number, and $\omega_0$ is the angular frequency.
The initial conditions of the wave packet are given by
\begin{eqnarray}
\rho &=& \rho_0 = 1 , \nonumber \\
p&=&p_0 = 0.9 , \nonumber \\
\VEC{v} &=& \VEC{0} , \nonumber \\
E_x &=& E_z = E_x^0 = E_y^0 = 0  , \nonumber \\
E_y &=& E_y^0 = E_0 \cos 2 \pi x e^{-x^2/2 \sigma^2}, \nonumber \\
B_x &=& B_y = B_x^0 = B_y^0 = 0  , \nonumber \\
B_z &=& B_z^0 = E_0 \cos \left (2 \pi x + \frac{\pi}{2} \right ) 
e^{-x^2/2 \sigma^2}, \nonumber \\
J_x &=& J_y = J_x^0 = J_y^0 = 0 , \nonumber \\
J_z &=& J_z^0 = 2^{3/2} \pi E_0 \cos 2 \pi x e^{-x^2/2 \sigma^2}.
\label{inipurwvpk}
\end{eqnarray}
Through this paper, we set $\sigma = 50$ and $E_0 = 0.01$.
We show the results of the wave packet propagation in Fig. \ref{fig2}.
We clearly found the shift of the wave packet from the initial position with 
$\Delta x = 4.2$ and $\Delta x = 7$ at $t=3$ and $t=5$, respectively. This
means that the infinite simple wave packet propagates with the group velocity
$v_{\rm g} = 1.4$. This confirms the results of \citet{koide08}.
It is also shown that the wave packet damps very quickly.
(Note that the scales of the vertical axes of the panels in Fig. \ref{fig2}
are quite different.)

\subsection{Propagation of a wave packet with a head edge}

\begin{figure}
\includegraphics{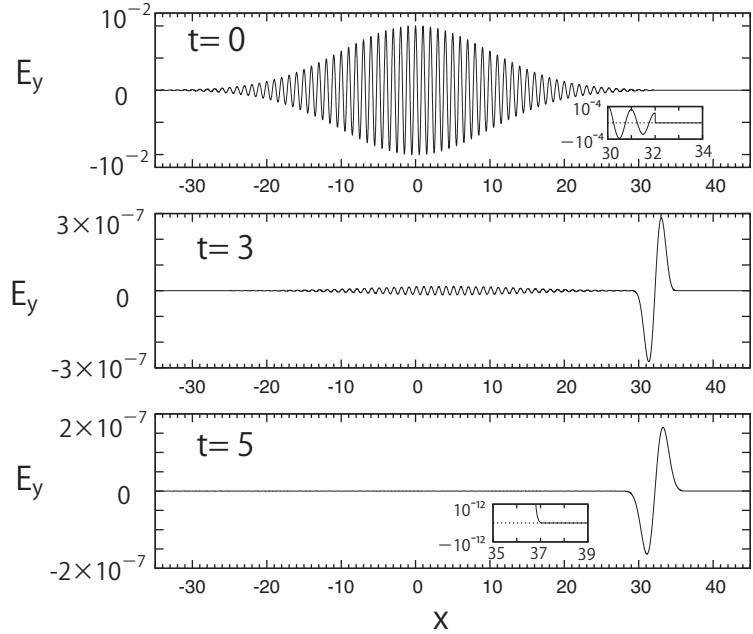}
\caption{
Simulation of a wave packet with a head edge.
The small boxes in the top and bottom panels show the zoom-up of 
the profiles of the
head edge of the wave packet in the range $30 \leq x \leq 34$,
$ -10^{-4} \leq E_y \leq 10^{-4}$ and $35 \le x \le 39$,
$-10^{-12} \leq E_y \le 10^{-12}$ at $t=0$ and $t=5$, respectively.
The latter small panel shows that the head edge locates at $x=37$ at $t=5$.
\label{fig3}}
\end{figure}

We consider a more realistic situation of information transport.
In this situation, we can not use the infinite length wave packet
at least without the head of the wave packet,
because it takes infinite time to produce such a wave packet.
Figure \ref{fig3} shows the wave packet propagation which
has the edge at the head $x=32$, $t=0$. 
The initial condition is given by multiplying the factor
\begin{equation}
f(x) = \left \{ 
\begin{array}{cl}
1 \\
0   & ( x > 32)  
\end{array} \right .  ,
\end{equation}
to each initial perturbation variable of the electromagnetic wave packet
expressed by Eq. (\ref{inipurwvpk}) as, 
\begin{equation}
E_i (x) = f(x) E_i^0(x), \  B_i (x) = f(x) B_i^0(x), \ 
{\rm and} \  J_i (x) = f(x) J_i^0(x).
\label{cutoff}
\end{equation}
Fig. \ref{fig3} clearly shows the bulk of the
wave packet propagates with the group velocity $v_{\rm g}=1.4$,
as shown in the case without the edge of the wave packet
(compare the peak of the wave packet part at $x=0$ and $x=4.2$
at $t=0$ and $t=3.0$, respectively).
On the other hand, at the front of the wave packet,
a relatively smoothed new structure shaped like combined upside-down
solitons (we name it ``head structure", hereafter) 
appears and becomes dominant compared
with the primary bulk part of the wave packet, because the head
structure damps more slowly than the bulk of the packet does
with a short characteristic wave length $\lambda_0 = 2 \pi/k_0 = 1$.
The front of the head structure propagates with (less than but almost) 
the light speed. It should be emphasized that the head structure 
becomes dominant before the bulk part
of the wave packet reaches to the initial position of the
wave packet front edge.
It is noted that even if the main part of the wave packet has an information,
we must begin with creating the front of the wave,
and then the velocity
of the information transport never exceeds the head velocity of the wave,
i.e., the light speed.
It is clearly shown that the wave packet can't transport
the information with the group velocity (which is larger 
than the light speed in the present case), and then the group velocity 
is not regarded as the velocity of the information propagation.
As shown in Section \ref{appendvwf}, the velocity limit of the 
information propagation is given by a variable called the head velocity.
Now, the first type of the causality problem 
of the standard resistive RMHD equations, about the information transportation
to the absolute past, is solved.

\section{Spontaneous decrease in entropy of wave packet \label{sec4}}

We consider the second type of the causality problem in the standard
resistive RMHD, that is, the suggested strange process of
spontaneous decrease in the entropy of the electromagnetic wave packet.
Before we calculate the wave packet propagation in the new coordinates 
frame $O'-x'y'z'$, where the electromagnetic energy of the wave packet 
is suggested to increase spontaneously, we have to 
consider the initial condition of the wave packet. When we use the approximation of 
the very narrow wave-number spectrum of the wave packet \citep{koide08}, 
the perturbation variables of the wave packet, for example, the transverse 
component of the electric field, $E_y$ is given by 
\begin{equation}
E_y(x,t) \propto \frac{1}{\sqrt{1+iDt/\sigma^2}}
\exp \left [ - \frac{1}{2 \sigma^2} \frac{(x-v_{\rm g} t)^2}{1+iDt/\sigma^2} \right ]
e^{-ik_0 x -i\omega_0 t}  ,
\end{equation}
where $\omega_0 = \omega(k_0)$, $v_{\rm g} = (\partial \omega/\partial k)_{k=k_0}$, 
and $D = (\partial^2 \omega/\partial k^2)_{k=k_0}$ \citep{koide08}.
As shown in Appendix \ref{appenda}, the profile of the initial condition
(Eq. (\ref{apinnf}))
in the new coordinates frame $O'-x'y'z'$ monotonically decreases
with respect to the coordinate $x'$. This is because the point at $x'<0$, $t'=0$
corresponds to the past of the original coordinates frame, $t<0$.
In the past, the amplitude of the  wave packet is (much) larger than
the initial profile in the plasma rest frame.
This electromagnetic wave with the monotonically decreasing profile
in the frame $O'-x'y'z'$ cannot be used as the transporter of information.
This is because, in the limit of $x' \longrightarrow - \infty$, 
the amplitude of $E'_y$ 
observed in the frame $O'-x'y'z'$ is approximately given by
$|E'_y| \sim \exp[-(\gamma_0^3 v_0^3 D^2/(\eta \sigma^2)) x']$, 
from Eq. (\ref{eqnol}), which is significantly large over an infinite region.
Here, $v_0$ is the relative velocity of the frame $O'-x'y'z'$ and
the frame $O-xyz$ and $\gamma_0 = (1 - v_0^2)^{-1/2}$ 
(see Eqs. (\ref{lorentztransf}) and (\ref{lorentztransfi})).
We cannot create such electromagnetic wave from a start source point of
information in finite time and with finite energy.
Then, to perform a simulation of the propagation of the signal 
with the wave packet,
we have to introduce an edge at the rear of the wave packet. 

\begin{figure}
\includegraphics{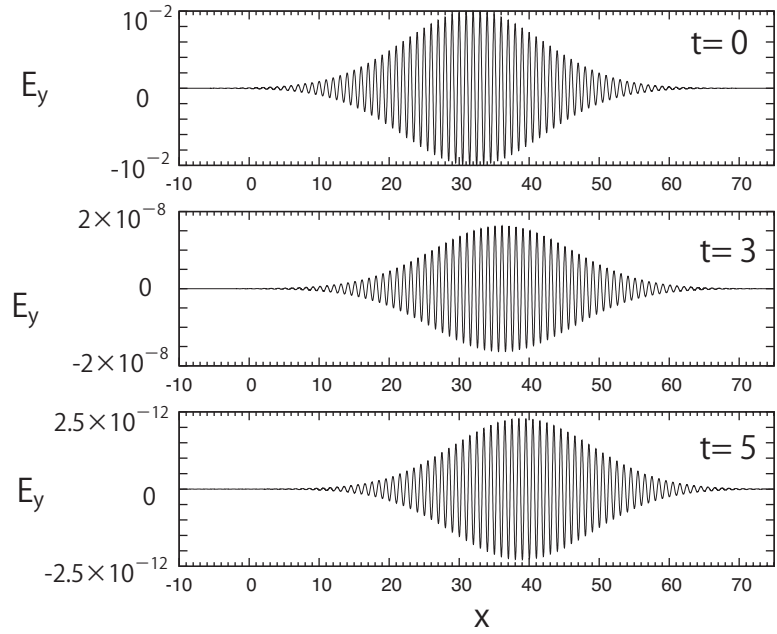}
\caption{
Simulation of propagation of wave packet with a very narrow Gaussian Fourier
spectrum. The initial condition of this wave packet shifts toward
the right direction by $\Delta x =32$ from the calculation of
Fig. \ref{fig1}.
\label{fig4}}
\end{figure}

To avoid the Lorentz transformation at $-x'\gg 1$ of the new coordinates frame,
we shift the wave packet with $\Delta x =32$ toward the left side at $t=0$
(Fig. \ref{fig4}). We also confirmed the primary wave packet propagates
with the group velocity $v_{\rm g} =1.4$.
We set the edge of the wave packet at the region $-6 \leq x \leq 0$, 
which is the relaxation zone of the step function at the wave packet edge.

\begin{figure}
\includegraphics{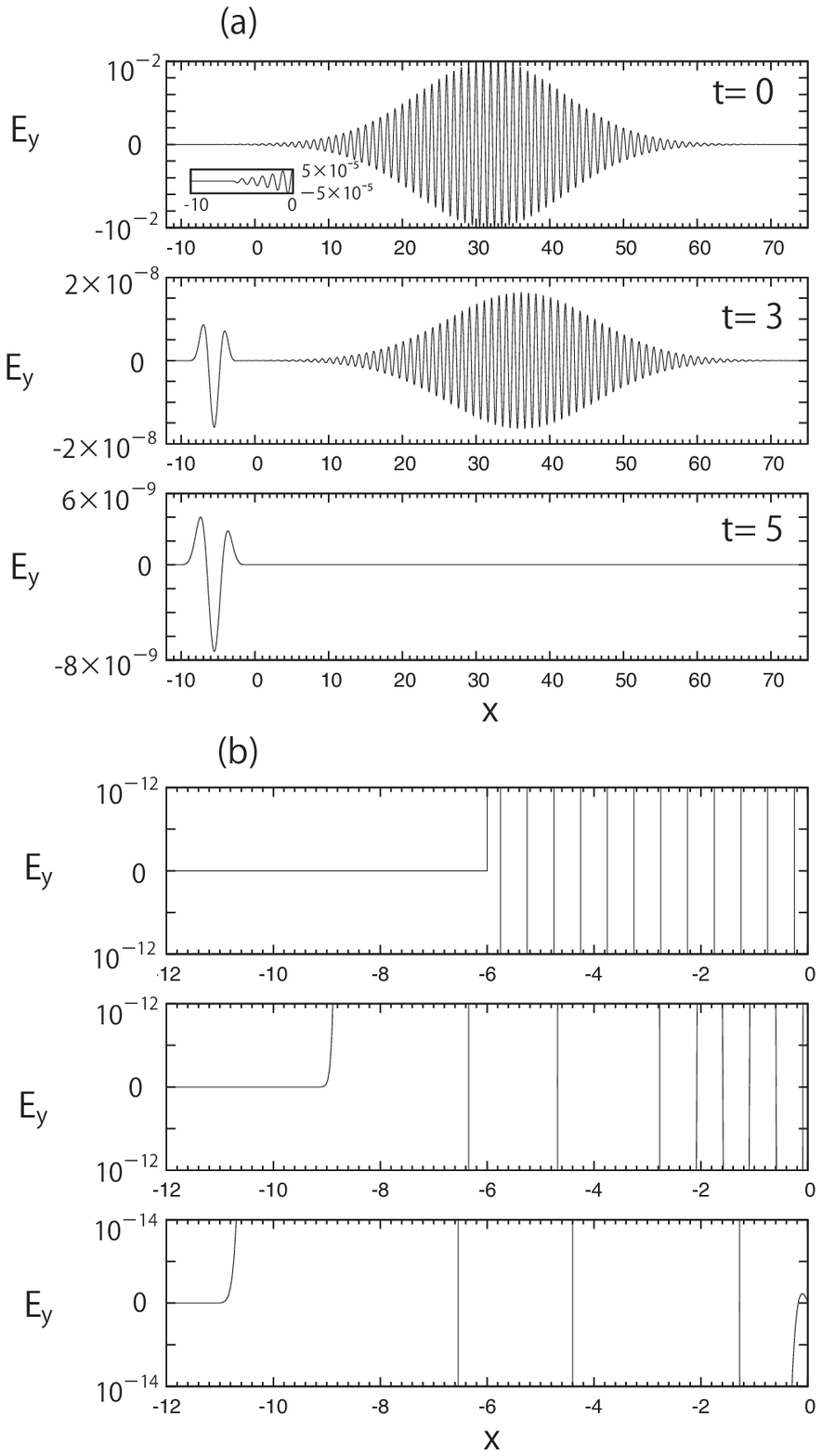}
\caption{
\label{fig5}}
Simulation of propagation of a wave packet with a back edge in the
plasma rest frame $O-xyz$.
(a) Whole profiles of the wave packet. (b) Profiles of the wave packet
around the rear edge.
The small box in the top panel shows the rear edge of the wave
packet in the range $-10 \leq x \leq 0$, 
$-5 \times 10^{-5} \leq E_y \leq 5 \times 10^{-5}$.
\end{figure}

\begin{figure}
\includegraphics{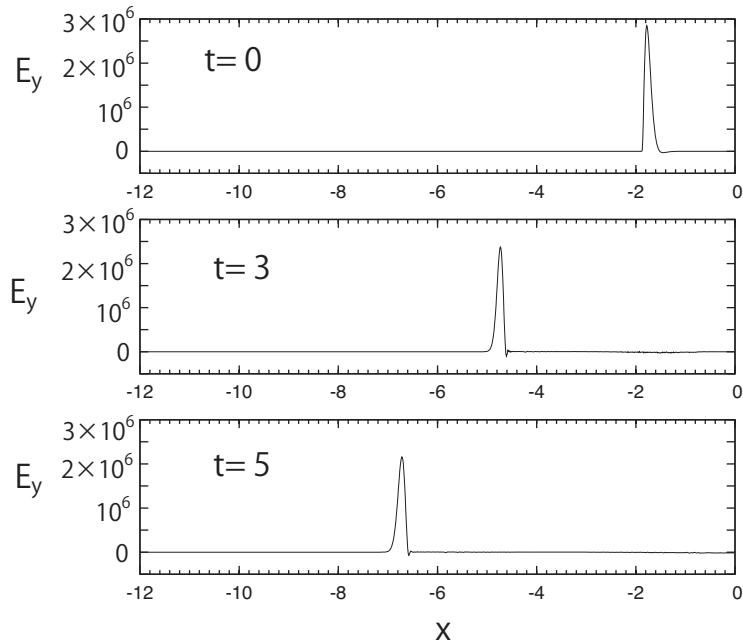}
\caption{
Simulation of propagation of a wave packet in the coordinates
frame where the plasma flows toward the left direction 
with $v_0>1/v_{\rm g}$. This coordinates frame corresponds to
the new frame $O'-x'y'z'$.
\label{fig6}}
\end{figure}

The initial condition is given by multiplying the following factor
to each initial variable of the electromagnetic wave packet as shown in
Eq. (\ref{cutoff});
\begin{equation}
f(x) = \left \{ \begin{array}{ccl} 
0  & & (x \leq -l) \\
1- \left ( \frac{x}{l} \right )^2 e^{(x^2-l^2)/2\sigma^2} 
& & (-l < 0 < 0) \\
1  & & (x \geq 0 )
\end{array} \right .
\end{equation}
where $l=6$ through this paper.
As the same of the front edge of the wave packet edge, the relatively smoothed
structure appears around the wave packet rear edge and becomes dominant
compared to the bulk wave packet (Fig. \ref{fig5}(a)).
Fig. \ref{fig5} (b) zooms up the edge of the wave packet and shows
that the left edge propagates with the light speed.
In this case, we can set the initial condition of the electromagnetic
wave as shown in the first panel of  Fig. \ref{fig6}.
Because the left of the $x'$-axis of the new coordinates frame
corresponds to the future of the primary coordinates frame
$O'-x'y'z'$, the bulk part of the wave becomes too small to see.
Only the rear edge part of the wave can be seen.
The edge part propagates toward the left direction with the light speed
and damps (Fig. \ref{fig6}).  This is reasonable because
the edge part propagates within the light speed in the plasma rest frame
(Fig. \ref{fig5}).
Then, the second type of expected strange phenomenon of the causality break-down
never happens.

\section{Wave head velocity as velocity limit of information propagation 
\label{appendvwf}}

In addition to numerical results with the electromagnetic wave packet, 
we theoretically confirm that 
the group velocity does not give the limit of the propagation
velocities of information and energy in this section. 
With respect to the information
velocity, we have confirmed numerically in Section \ref{sec3} 
that the group velocity presents the propagation
velocity of the wave packet without front or rear edge
whose Fourier component is a Gaussian function
with a narrow spectrum. 
While most part of the initial profile of the wave packet 
is concentrated in the region with the length of several $\sigma$, 
the amplitude
of the profile is finite over the whole range, $-\infty < x < \infty$.
In the inertia frame, whose relative velocity to the plasma rest frame
exceeds the inverse of the group velocity ($v_0 > 1/v_{\rm g}$), 
the wave packet does not transport an information 
because its shape becomes a monotonic function and has no peak
(Appendix \ref{appenda}). Then, we recognize such a wave packet which spread
over the whole region $-\infty < x < \infty$ is not
accepted as an object of information transporter in relativity. 
Concerning the energy propagation
velocity, in the first place, we note that we cannot define the 
velocity of energy propagation 
because energy is a temporal component of 4-momentum and we cannot consider
the transportation of energy only. In relativity, we have to consider
transportation of 4-momentum, not only the energy component. 
In fact, the group velocity $v_{\rm g} = \partial \omega/\partial k$
is not a scalar, much less 4-vector, 
and then the superluminal group velocity $v_{\rm g} > c$
is not an invariant statement in any inertial coordinates frame.
This contradicts to relativity.
Then, we understand the 
propagation velocity of energy has no physical meaning in relativity. 

Instead of the propagation velocity of the wave packet
without the edge (i.e., the group velocity), 
we can introduce the ``head velocity",
\begin{equation}
v_{\rm h} \equiv \lim_{\omega \longrightarrow \infty} \frac{\omega}{k} ,
\label{defvh}
\end{equation}
as the velocity limit of the propagation of non-zero fields.
The head velocity $v_{\rm h}$ is a scalar
if it is the light speed in a certain coordinates frame, and
it is less than the light speed in any inertia frame if it is 
in a certain frame (Appendix \ref{appendvheadinvariant}).
It has been already shown that this head velocity represents the information
velocity limit in the dielectrics 
(see section 7.11 of \citet{jackson62}, and references therein).

\begin{figure}
\includegraphics[scale=0.75]{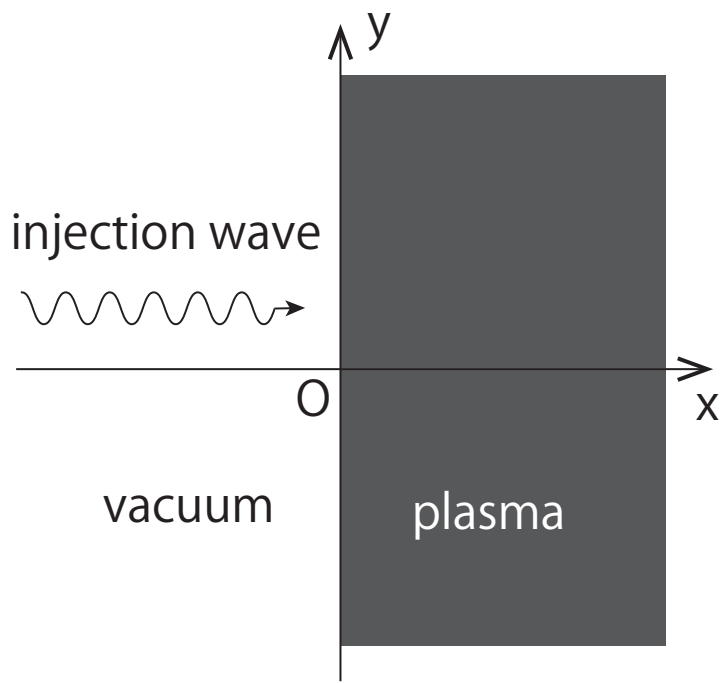}
\caption{Injection of electromagnetic simple plane wave
from the vacuum to the semi-infinite uniform plasma at $x \ge 0$.
\label{fig7}}
\end{figure}

\begin{figure}
\includegraphics[scale=0.8]{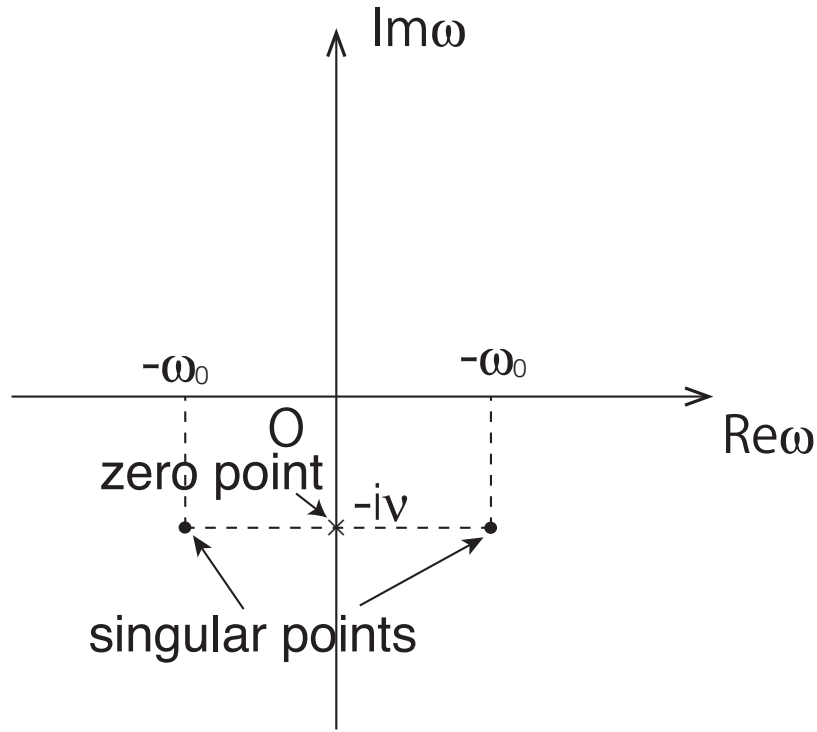}
\caption{Singular and zero points of $A(\omega)$ on
the $\omega$ complex plane.
\label{fig8}}
\end{figure}

Here, we prove that the head velocity of the wave
provides the velocity limit of information propagation also in a plasma.
This proof confirms the results of Section \ref{sec3} with respect
to the first problem of causality shown in Section \ref{sec2}.
To investigate the propagation speed of the electromagnetic plane wave
front, we consider an injection of the electromagnetic plane wave
normally incident from the vacuum on the semi-infinite uniform plasma filling
in the region $x>0$, whose boundary locates at $x=0$ at $t=0$ 
(Fig. \ref{fig7}).
We assume the plasma has uniform resistivity $\eta$, and the dispersion
relation of the electromagnetic wave is presented by Eq. 
(\ref{dispersion_relation}).
The index of refraction $n(\omega)$ is given by
\begin{equation}
n(\omega) = \frac{k(\omega)}{\omega} = \sqrt{1 + \frac{i}{\eta \omega}}.
\end{equation}
According to Eqs. (7.124) and (7.125) of \citet{jackson62}, the
amplitude of the electric field of the plane wave for $x>0$ is given by
\begin{equation}
u(x,t) = \int_{-\infty}^{\infty} \frac{2}{1+n(\omega)}
A(\omega) e^{ik(\omega) x - i \omega t} d \omega ,
\label{appenduxt}
\end{equation}
where 
\begin{equation}
A(\omega) = \frac{1}{2 \pi} \int_{-\infty}^{\infty}
u_{\rm i} (0,t) e^{i\omega t} dt
\end{equation}
is the Fourier transform of the real incident electric field
$u_{\rm i} (x,t)$.
Suppose that the incident wave has a well-defined front edge that
reaches $x=0$ at $t=0$. Then, the incident electric field $u_{\rm i}(x,t)$
evaluated just outside the plasma at $x=0^-$ has the condition,
\begin{equation}
u_{\rm i}(0,t)=0 \verb!   ! (t<0).
\label{appendui0t}
\end{equation}
This condition, together with certain mathematical requirements that
are ``physically reasonable", is both necessary and sufficient to
assure that $A(\omega)$ is analytic in the upper half of the
complex $\omega$ plane. In this situation, the
``physically reasonable" requirement is that 
$u_{\rm i}(0,t) \longrightarrow 0$ as $t \longrightarrow \infty$.
This is true at least for the wave packet, 
while for semi-infinite continuous wave
train at $0 < t < \infty$, the physical reasonable condition
is not satisfied\footnote{It means that the initial 
condition of the rear part of the wave very far form the front edge
determines the propagation of the wave front.
This case is out of our scope.}.
For an example, we consider a simple case of
\begin{equation}
u_{\rm i}(0,t) = \left \{ 
\begin{array}{cl}
E_0 e^{-\nu t} \cos \omega_0 t & (t \ge 0) \\
0   & ( t < 0)  \end{array} \right .  ,
\end{equation}
where $E_0$, $\nu$, and $\omega_0$ are the positive variables
(when $\nu =0$, it is not the case of ``physically reasonable").
Then, we have 
\begin{equation}
A(\omega) = \frac{i E_0}{2 \pi} 
\frac{\omega + i \nu}{(\omega + i \nu)^2 - \omega_0^2} .
\end{equation}
It has singular points at $\omega = \pm \omega_0 - i \nu$
and a zero point at $\omega = - i \nu$ (Fig. \ref{fig8}). It clearly shows that
there is no singular point in the upper half of the $\omega$ plane.
In the limit of $|\omega| \longrightarrow \infty$, we have
$A \longrightarrow \frac{iE_0}{2 \pi \omega}$, then
$|A| \leq \frac{E_0}{2 \pi |\omega |}$. This inequality can be
generalized for an arbitrary incident dumping wave train like
a wave packet, if we take the constant $E_0$ large enough.

\begin{figure}
\includegraphics[scale=0.8]{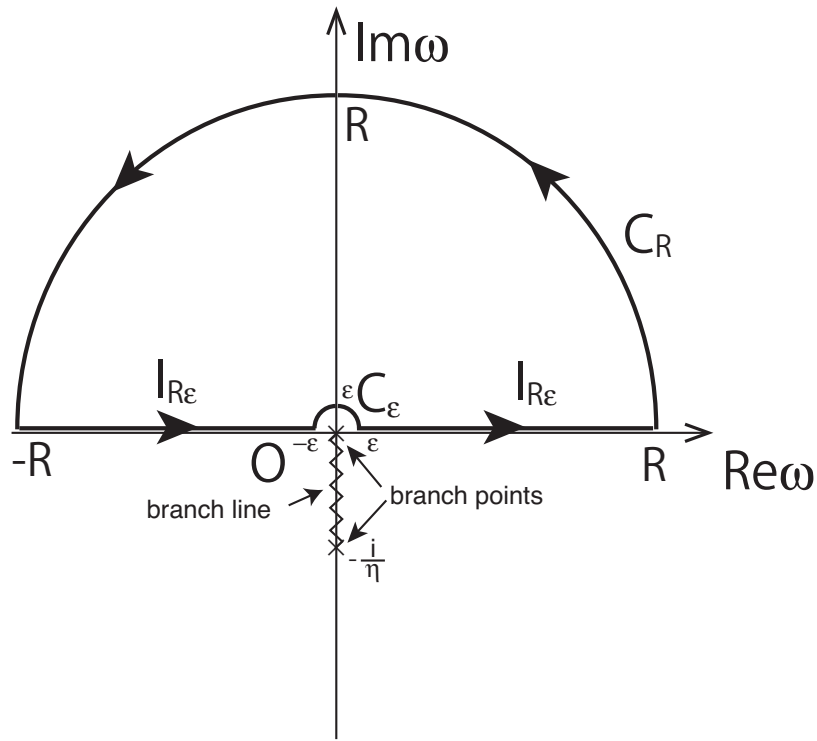}
\caption{Branch points and a branch line of $k(\omega)$ on
the complex $\omega$ plane and the path of the contour
integral for the amplitude of the electric field (Eq. (\ref{appenduxt})).
\label{fig9}}
\end{figure}

We formally evaluate the amplitude presented by 
Eq. (\ref{appenduxt}) by contour integration. 
For a plasma case, the path of the contour integral includes the branch point
of $k(\omega) = \omega (\omega + \frac{i}{\eta})$ at $\omega = 0$.
To avoid the multi-variable definition of $k(\omega)$ around $\omega = 0$,
we have to modify the integral path near the branch point $\omega = 0$
so as not to cross the branch line between $\omega = 0$ and 
$\omega = -\frac{i}{\eta}$ as shown in Fig \ref{fig9}.
Here, the radius of the half circle $C_\epsilon$ around $\omega =0$ is
infinitesimally small ($\epsilon \longrightarrow 0$).
With the definition of the head velocity $v_{\rm h}$ (Eq. (\ref{defvh})), 
the argument of the exponent
in Eq. (\ref{appenduxt}) becomes $i \omega ( x -v_{\rm h} t)$ for large 
$\omega$,
and the closing of the contour can be done in the upper half plane
for $x>v_{\rm h} t$, and the lower half plane for $x<v_{\rm h} t$. 
With respect to the case of $x>v_{\rm h} t$, this is because 
we have the inequality with respect 
to the integration along
the large upper half circle $C_R: z=R e^{i \theta}$ ($0 \leq \theta \le \pi$),
where $R$ is the radius of the half circle $C_R$,
\begin{equation}
\left | \int_{C_R} \frac{2}{1 + n(\omega)} 
A(\omega) e^{ik(\omega)x - i \omega t} d \omega \right |
\le E_0 \frac{|1 - e^{-(x-v_{\rm h} t)R}|}{(x-v_{\rm h} t)R} ,
\end{equation}
and then the integral along $C_R$ vanishes in the limit of 
$R \longrightarrow \infty$.
In Fig. \ref{fig9}, we can see the integration contour 
consists of $C_\epsilon$, $C_R$, and $I_{\epsilon,R}$
for the case of $x > v_{\rm h} t$.
With the analytic nature of $n(\omega)$ and $A(\omega)$ in the upper half
of the $\omega$ Riemann plain, the whole integrand in the right-hand side 
of Eq. (\ref{appenduxt})
is analytic there, and the Cauchy's theorem guarantees 
that the integral vanishes.
We have thus proved that
\begin{equation}
u(x,t) = 0 \verb!   ! (x > v_{\rm h} t)
\label{appenduxt0}
\end{equation}
provided that $A(\omega)$ and $n(\omega)$ are analytic for
$\Im \omega > 0$ and $n(\omega) \longrightarrow 1/v_{\rm h}$ for
$|\omega| \longrightarrow \infty$.
Equation (\ref{appenduxt0}), together with Eq. (\ref{appendui0t}), establishes
that no signal propagates with a velocity greater than the head velocity,
in the plasma. This confirms the results of Section \ref{sec3} about
the velocity limit of the electromagnetic wave packet, but not those of
Section \ref{sec4} with respect to the causality problem of
decrease in the entropy.
In the plasma, the head velocity becomes the light velocity, $v_{\rm h}=1$.

\section{Usage of standard resistive RMHD equations \label{secdiscuss}}

Now, we understand the standard resistive RMHD equations have no causality
problem, although the group velocity of the electromagnetic wave calculated 
by these equations exceeds the light speed.
We showed that the head velocity of the wave determines the information
propagation velocity limit as shown in the previous section.
Then, we are free from the restraint of 
the superluminal group velocity in the standard 
resistive RMHD equations. Here, we consider the validity and limit of the 
standard resistive RMHD equations based on a more general point of
view using the generalized RMHD equations.
We already evaluated each term of the generalized general
relativistic MHD (GRMHD) for 
the plasmas around several astrophysical black holes \cite{koide10}, 
where we compare these terms with the $\VEC{V} \times \VEC{B}$ term.
We compare the terms beyond the standard resistive RMHD equations with 
the resistive terms in generalized relativistic Ohm's law, first,
\begin{eqnarray}
\frac{1}{ne} \frac{\partial}{\partial t} \left (
\frac{\mu h}{ne} \VEC{J}^\dagger \right )
&=& - \frac{1}{ne} \nabla \cdot \VEC{K}
+ \frac{1}{2ne} \nabla (\Delta \mu p - \Delta p) \nonumber \\
&& + \left ( \gamma - \frac{\Delta \mu \rho_{\rm e}}{ne} \right ) \VEC{E}
+ \left ( \VEC{U} - \frac{\Delta \mu}{ne} \VEC{J} \right ) \times \VEC{B}
- \eta [\VEC{J} - \rho_{\rm e}' (1 + \Theta) \VEC{U}],
\label{genrelohm}
\end{eqnarray}
in flat space-time, which comes from Eq. (63) of \citet{koide10}.
Here, $n$ is the particle number density $n \equiv \rho /m$,
$\frac{\mu h}{ne} \VEC{J}^\dagger$ is the momentum density of the electric current,
$\VEC{K}$ is the stress tensor of charge and current, $m$ is the characteristic
mass of a particle $m \equiv m_+ + m_-$, $\mu$ is the normalized reduced mass,
$\mu = m_+ m_-/(m_+ + m_-)^2$, $\Delta \mu$ is the normalized mass
difference $\Delta \mu = (m_+ - m_-)/(m_+ + m_-)$, $\Delta p = p_+ - p_-$, and $\Theta$ is the
thermal energy exchange rate from the negatively charged fluid and the positively 
fluid, where $m_\pm$ is the mass of positively/negatively charged particle, 
$p_\pm$ is the pressure of positively/negatively fluid in two-fluid model.
The left-hand side of Eq. (\ref{genrelohm}) presents the inertia of current,
the first term of the right-hand side shows the transport of electric current
momentum, the second term indicates the thermoelectromotive force,
and the term $\eta \rho_{\rm e}' \Theta \VEC{U}$ in the last term
expresses the redistribution of the thermal energy due to the friction 
between the two fluids.

We evaluate these terms beyond the standard relativistic Ohm's law except
for the last term about thermal energy redistribution of friction.
\begin{itemize}
\item Inertia of electric current: the left-hand side of Eq. (\ref{genrelohm}),
\begin{equation}
\frac{\rm (inertia \; of \; current)}{\rm (resistive \; term)} =
\frac{1}{\eta J} \frac{1}{ne} \frac{\partial}{\partial t} \left ( 
\frac{\mu h}{ne} J \right ) \sim \frac{1}{\eta J} \frac{\omega \mu h J}{(ne)^2}
= \mu \frac{\omega}{\nu_{\rm e*}},
\end{equation}
where $\omega$ is a characteristic frequency of phenomena,
$J$ is a characteristic value of the current density, and $\nu_{\rm e*}$ is
the Coulomb collision frequency between electron and positively charged 
particle (*=i: ion, *=e: positron).
Here, we used the relations, $\eta = m \nu_{\rm e*}/(ne^2) \Rightarrow 
h \nu_{e*}/(ne)^2$, $\omega_{\rm p} = ne/\sqrt{\mu h}$, and 
$\omega_{\rm c} = enB/h$. 
The Coulomb collision frequencies of electron-positron ($\nu_{\rm ee}$) 
and electron-ion ($\nu_{\rm ei}$) are evaluated by
\begin{eqnarray}
\nu_{\rm ee} &=& 1.6 \times 10^{10} \left ( \frac{T_{\rm e}}{e} \right )^{-3/2}
\left ( \frac{n_{\rm e}}{10^{20}} \right ) \left [ {\rm s}^{-1} \right ] , \\
\nu_{\rm ei} &=& 6.3 \times 10^{9} \left ( \frac{T_{\rm e}}{e} \right )^{-3/2}
\left ( \frac{n_{\rm e}}{10^{20}} \right )  \left [ {\rm s}^{-1} \right ] ,
\end{eqnarray}
where unit of $T_{\rm e}/e$ is eV and unit of $n_{\rm e}$ is $\rm m^{-3}$
(Appendix in \citet{miyamoto87}).
Then, these collision frequency are comparable,
$\nu_{\rm ee} \sim 2.5 \nu_{\rm ei}$.
\item Thermoelectromotive force: the second term of right-hand side of Eq. 
(\ref{genrelohm}),
\begin{eqnarray}
\frac{\rm (thermoelectromotive \; force)}{\rm (resistive \; term)}
&=& 
\frac{1}{\eta J} \frac{1}{2ne} |\nabla (\Delta \mu p - \Delta p)| \nonumber \\
&& \sim \frac{1}{\eta J} \frac{1}{2ne} \frac{1}{L} |\Delta \mu p - \Delta p|
\sim \frac{\Delta \mu p - \Delta p}{2h} 
\frac{\mu \omega_{\rm p}^2}{\nu_{\rm e*} \omega_{\rm c}} ,
\end{eqnarray}
where $L$ is a characteristic length of the phenomena.
\item Hall effect: the third term of right-hand side of Eq. (\ref{genrelohm}),
\begin{equation}
\frac{\rm (Hall \; term)}{\rm (resistive \; term)} = \frac{1}{\eta J}
\frac{\Delta \mu}{ne} |\VEC{J} \times \VEC{B} | \sim 
\frac{1}{\eta J} \frac{\Delta \mu}{ne} JB 
= \Delta \mu \frac{\omega_{\rm c}}{\nu_{\rm e*}} ,
\end{equation}
where $B$ is a characteristic value of magnetic field.
\end{itemize}

In the generalized RMHD equations, the stress tensor is given by
\begin{equation}
T^{ij} = p \delta^{ij} + h U^i U^j + \frac{2 \mu \Delta h}{ne} (U^i J^j + J^i U^j )
+ \frac{\mu h^\ddagger}{(ne)^2} J^i J^j 
+ \left ( \frac{B^2}{2} + \frac{E^2}{2} \right ) \delta^{ij}
-B_i B_j -E_i E_j .
\label{stress_tensor}
\end{equation}
The momentum stress tensor of electric current is shown by the third and forth
terms of the right-hand side of Eq. (\ref{stress_tensor}).
The terms are evaluated by comparison with the kinetic stress tensor $h U^i U^j$ as
\begin{equation}
\frac{\rm (current \; momentum \; tensor)}{\rm (hydrodynamic \; stress 
\; tensor)} \sim
\frac{2 \mu \Delta h U J }{U^2 n e h} \leq 2 \mu \frac{J}{U n e}
\sim 2 \mu \frac{\eta}{UL} \frac{B}{\eta n e} 
= \frac{2 \omega \omega_{\rm c}}{\omega_{p}^2} .
\end{equation}
The magnetic Reynolds number $S_{\rm M}$ is defined by
\begin{equation}
S_{\rm M} = \frac{U L}{\eta} =
\frac{(ne)^2 UL}{h \nu_{\rm e*}} = \frac{\mu \omega_{\rm p}^2 U L}{\nu_{\rm e*}}
\leq \frac{\mu \omega_{\rm p}^2 L}{\nu_{\rm e*}} = 
 \mu \frac{\omega_{\rm p}^2}{\nu _{\rm e*} \omega}, 
\end{equation}
where $U$ is a characteristic values of 4-velocity $|\VEC{U}|$.
Table 1 shows the values of these ratios for plasmas around several types
of astrophysical black holes with the mass $M_{\rm BH}$.
Here, we assume that the plasma is electron-ion plasmas.
We use the wavelength of the fastest-growing magnetorotational instability 
(MRI), $L_{\rm CS}$, which causes the current sheet in the plasma disk
as the characteristic length of phenomena around the black holes $L$.
It is noted that the wavelength of the fastest-growing mode gives 
the maximum thickness of the current. Because in the linear stage, the magnetic 
field lines are bent with the scale of the wavelength and the current sheet
is formed with the scale. In the nonlinear stage, the current sheet
is pinched until the magnetohydrodynamic equilibrium between
the pressure gradient and Lorentz force is achieved. 
It was shown in numerical MHD calculations (for examples, 
\cite{hawley92,sano01}).
The wavelength of MRI at maximum growth rate is given by $L_{\rm CS} \sim
4 \sqrt{ \frac{2}{3}} \frac{v_{\rm A}}{\Omega}$, where $\Omega$ is the
angular velocity of the disk around the black hole, $\Omega \approx
\sqrt{G M_{\rm BH}/r^3} > r_{\rm S} \equiv 2 G M_{\rm BH}$ and $v_{\rm A}$ is
the Alfven velocity, $v_{\rm A} = \sqrt{B^2/\rho}$ ($G$ is the gravitational
constant).
It shows that the magnetic Reynolds number is much larger than the unity 
and the 
resistivity is negligible for the scale of current sheet (see also page 1468,
right column of \citet{koide10}). With respect to Ohm's laws, we can neglect
the inertia of current in all cases. Hall effect is negligible for supermassive
black hole cases, while not for gamma-ray bursts (GRBs) and black hole (BH) 
X-ray binaries (including stellar mass black holes).
On the other hand, the thermoelectromotive force would become 
significant compared to the resistivity in all situations. 
Because of the huge values of magnetic Reynolds number
in Table 1, the electromotive force due to resistivity is very small 
compared to the ideal MHD electromotive force and then Hall effect
is the same.
With respect to the momentum transportation of current, 
it is negligible for all cases. In conclusion, we can use standard resistive RMHD
equations with additional terms of thermoelectromotive force 
(and Hall term for GRB and BH X-ray binary.)

When we consider the pair plasma, the normalized reduced mass and normalized
mass difference become $\mu = 1/4$ and $\Delta \mu = 0$ from
$\mu = m_{\rm e}/m_{\rm i} \sim 1/1800$ and $\Delta \mu = 0$
in the electron-ion plasma, respectively.
Hall effect disappears 
and other special effects of generalized RMHD may be greater than the
ratios of normal plasma by $m_{\rm i}/m_{\rm e} \sim 1800$. The conclusion 
with respect to the validity of the standard resistive RMHD equations 
is not changed.

\section{Summary \label{sec5}}

We investigated whether the standard resistive RMHD equations have the causality
problem or not using the direct numerical calculations with the equations.
There were two kinds of the hypothetical causality problem: 
(i) information transportation
to the absolute past and (ii) spontaneous decrease in entropy. The numerical
calculations showed that such phenomena do not appear even in cases
where such phenomena are expected to happen.
In the calculations, we used the electromagnetic wave packet with
the characteristic length $\sigma = 50$, characteristic wave number
$k_0 =2 \pi$ in the plasma with resistivity $\eta = 1/(2^{3/2} \pi)$.
These results suggest that the ``group velocity" does not have
physical meaning, such as the propagation velocity of information.
In fact, with respect to the first causality problem (i),
we confirmed that the head velocity of the wave represents
the velocity limit of the information propagation and the head velocity
is the light speed in the plasma.

\begin{figure}
\includegraphics{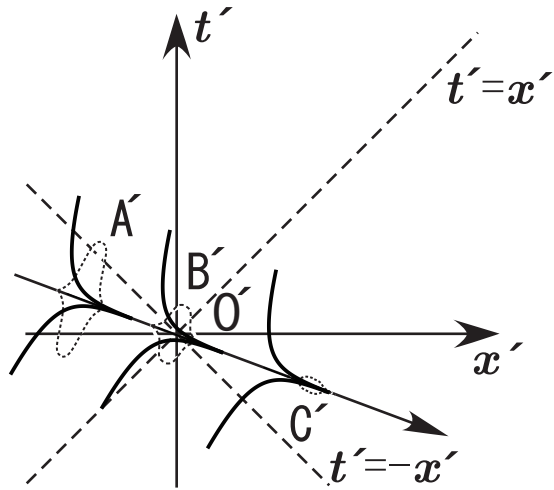}
\caption{Corrected image of the panel (b)
of Fig. \ref{fig1}. In the frame $O'-x'y'z'$, the profile
of the electromagnetic wave becomes monotonic and transports
no signal because it has no peak.
\label{fig10}}
\end{figure}

With respect to the second problem of causality (ii),
the calculation results suggest that, in general, acausal phenomena with
a wave packet never happen because the wave packet without an edge in the 
plasma rest frame never be transformed to the wave packet
with a peak by the Lorentz transformation with
the relative velocity $v_0 > 1/v_{\rm g}$, as shown in Fig. \ref{fig10} 
(Appendix \ref{appenda}), and the wave packet with an edge induces
the (pair-soliton-like) head structure at the 
edge of the wave packet independently (or weakly dependent) 
on the bulk wave packet.
The former shows that the concept of a wave packet is not consistent with
the frame-invariant concept of relativity.
Consequently, we understand that we have no causality problem in
standard resistive RMHD as suggested by discussion with the superluminal 
group velocity
of the electromagnetic wave \citep{koide08}. This conclusion gives the
guarantee of an usage of the standard resistive RMHD.
This problem was the (mentally) biggest obstacle for application of the
standard resistive RMHD equations to the global astrophysical phenomena. 
We expect that numerical simulations with the standard resistive RMHD
equations to be more popular for analysis of plasmas around
black holes, where the magnetic reconnection may play an important
role in the release of the energy stored in the plasma near
the black holes.

Incidentally, it is noted that in the plasma, whose plasma parameter is much
larger than unity, the group velocity is less than the light speed \citep{koide08}.
To include this effect correctly, we have to include the inertia effect
of the electron or electric charge/current. That is, we have to
use generalized RMHD equations to consider the high frequency
phenomena, where we cannot neglect the electron or current inertia,
while it is not necessary for the global, relatively slow phenomena around
the astrophysical black hole.

\begin{acknowledgments}
I am grateful to Mika Koide for her helpful comments on this paper.
I thank Takahiro Kudoh for the fruitful discussion.
This work was supported in part by the Science Research Fund of
the Japanese Ministry of Education, Culture, Sports, Science and Technology.
\end{acknowledgments}

\appendix

\section{Initial condition of wave packet in a case of hypothetical 
acausal effect 
\label{appenda}}

We show the profile of the electromagnetic wave packet propagating along the $x$ 
direction becomes monotonic when we observe it from the coordinates frame
to the relative velocity $v_0 > 1/v_{\rm g}$ to the rest frame of
the plasma, $O-xyz$. Here, in the frame $O-xyz$, we assume that the
Fourier spectrum of the wave packet is Gaussian,
\begin{equation}
F(k) = \frac{\sigma}{\sqrt{2 \pi}} e^{- \frac{\sigma^2}{2}(k-k_0)^2},
\end{equation}
where $\sigma$ corresponds to the characteristic length of the wave
packet and $k_0$ is the typical wave length of the wave packet.
Furthermore, we assume $\sigma$ is large enough and the Fourier spectrum
is significant only near $k \sim k_0$. We have the complex perturbation
variables of the electromagnetic wave packet as
\begin{equation}
E_y(x,t) \propto \frac{1}{\sqrt{1+iDt/\sigma^2}}
\exp \left [ - \frac{1}{2 \sigma^2} \frac{(x-v_{\rm g} t)^2}{1+iDt/\sigma^2} \right ]
e^{-ik_0 x -i\omega_0 t}  ,
\label{terminate}
\end{equation}
where $\omega_0 = \omega(k_0)$, $v_{\rm g} = (\partial \omega/\partial k)_{k=k_0}$, 
and $D = (\partial^2 \omega/\partial k^2)_{k=k_0}$ \citep{koide08}.
The physical solution is given by the real part of the complex perturbation, $E_y$.
The amplitude around any point $x=x_0$ at any time $t=t_0$ is given by,
\begin{eqnarray}
E_y(x,t) &\propto& \frac{1}{|\sqrt{1+iDt/\sigma^2}|}
\exp \left [\Re \left\{ - \frac{1}{2 \sigma^2} 
\frac{(x-v_{\rm g} t)^2}{1+iDt/\sigma^2} \right \} \right ]
e^{(\Im \omega_0)t}  , \\
& = & \frac{1}{\sqrt{1+(Dt/\sigma)^2}} \exp \left [ 
-\frac{1}{2 \sigma^2} \frac{(x - v_{\rm g})^2}{1+(Dt/\sigma)^2}
\right ] e^{(\Im \omega_0) t} .
\end{eqnarray}
The Lorentz transform from the original coordinates frame $O-xyz$ to
the new coordinates frame $O'-x'y'z'$ with the relative velocity
$\VEC{v}_0 = v_0 \hat{x}$,
\begin{equation}
t' = \gamma_0 (t - v_0 x), x' = \gamma_0 (x - v_0 t), y' = y, z' = z,
\label{lorentztransf}
\end{equation}
or
\begin{equation}
t = \gamma_0 (t' + v_0 x'), x = \gamma_0 (x' + v_0 t'), y = y', z = z',
\label{lorentztransfi}
\end{equation}
where $\gamma_0 = (1- v_0^2)^{-1/2}$.
Then, the amplitude profile of the initial electric field of the wave packet
observed by the new coordinates frame $O'-x'y'z'$ is
\begin{eqnarray}
|E_y'| &=& |\gamma_0 (E_y - v_0 B_z)| = \gamma_0 \left |1 - \frac{v_0 k}{\omega} \right |
|E_y| \\
&\propto& \frac{1}{[1+(D\gamma_0 v_0 x'/\sigma^2)^2]^{1/4}}
\exp \left [ 
- \frac{1}{2 \sigma^2} \frac{(1-v_{\rm g} v_0)^2 x'^2}{1+(D \gamma_0 v_0 x'/\sigma^2)^2}
\right ] e^{\gamma_0 v_0 (\Im \omega_0) x'}  ,
\label{apinnf}
\end{eqnarray}
where we assumed $v_0 \neq \omega_0/k_0$.
Here, we have the dispersion relation of the electromagnetic wave in the
plasma with the uniform resistivity $\eta$, given by 
Eq. (\ref{dispersion_relation}).
Then, we have
\begin{eqnarray}
\omega_0 &=& - \frac{i}{2 \eta} \pm \frac{1}{2 \eta} \sqrt{4 \eta^2 k^2 -1}, \\
v_{\rm g} &=& \frac{2 \eta k}{\sqrt{(2 \eta k)^2-1}},  \\
D &=& - \frac{1}{2 \eta^2} \left ( \frac{2 \eta}{\sqrt{(2 \eta k)^2-1}} \right )^3 < 0. 
\end{eqnarray}
Here, we have the relation,
\begin{equation}
\frac{D}{\eta} = - \frac{1}{2} \left ( \frac{v_{\rm g}}{\eta k}   \right )^3.
\end{equation}
We note that $|E_y|$ becomes infinite when $x' \longrightarrow - \infty$
because $\Re \omega_0 = -1/\eta < 0$, $D \neq 0$, and $v_0 \neq 0$.
This means that this wave packet keeps its finite profile only in the
original frame $O-xyz$.
In other coordinates frame $O'-x'y'z'$ with the finite relative velocity
to the rest frame of the plasma, $O-xyz$, the electromagnetic wave
can not be recognized as a wave packet anymore.
Then, we understand that the concept of ``wave packet" is not invariant item
with respect to the Lorentz transformation.

To show the monotonicity of the initial perturbation in the new coordinates
$O'-x'y'z'$, we evaluate the following variable proportional to the logarithm of the profile 
of the perturbation,
\begin{equation}
L = 2 \left ( \frac{D \gamma_0 v_0}{\sigma} \right )^2
\left [ - \frac{\gamma_0 v_0}{2 \eta} x' 
- \frac{(1-v_{\rm g} v_0)^2 \sigma^2}{2[\sigma^4+(D\gamma_0 v_0 x')^2]} x'^2
-\frac{1}{4} \log \left  \{ 
\sigma^4 + (\gamma_0 v_0 D x')^2 \right \}
\right ] .
\end{equation}
Using a new variable $\xi = - \frac{\gamma_0 v_0 D}{\sigma^2} x'$, we have
\begin{equation}
L =  \frac{(\gamma_0 v_0)^2 D}{\eta} \xi
+ (v_{\rm g} v_0 - 1)^2 \frac{\xi^2}{1 + \xi^2}
-\frac{1}{2}  \left ( \frac{D \gamma_0 v_0}{\sigma} \right )^2
\log \sigma^4 ( 1+\xi^2) .
\label{eqnol}
\end{equation}
We evaluate the slop of $L$,
\begin{equation}
\frac{\partial L}{\partial \xi} = 
\frac{D (\gamma_0 v_0)^2}{\eta} \left [
1 - \frac{\eta D}{\sigma^2} \frac{\xi}{1 + \xi^2}  \right ]
+ (v_{\rm g} v_0 - 1)^2 \frac{2 \xi^2}{(1 + \xi^2)^2}
\end{equation}
When $\xi$ is greater than zero, it is clear that $\partial L/\partial \xi$ 
is negative.  When $\xi$ is negative, $\partial L/\partial \xi$ is
negative if
\[
\frac{1}{\gamma_0^2} \frac{\partial L}{\partial \xi}
< -4 v_0^2 ( v_{\rm g}^2 -1 )^{3/2}
+ \left ( \frac{\sqrt{3}}{2} \right )^3 (v_{\rm g} v_0 -1)^2 (1 - v_0^2)
+ 2 v_0^2 ( v_{\rm g}^2 -1 )^3 \frac{4\eta^2}{\sigma^2}  < 0,
\]
where we use inequalities, $|2\xi/(1+\xi^2)| \leq 1$ and $|2\xi/(1+\xi^2)^2| \leq (\sqrt{3}/2)^3$.
When $v_0 > 1/v_{\rm g}$ and $v_{\rm g} > 1$, we have 
\[
(v_{\rm g} v_0 -1)^2 \left ( 1 - \frac{1}{v_0^2} \right) < 
\left ( v_{\rm g}^{2/3} - 1 \right )^3.
\]
and
\[
\frac{\left ( v_{\rm g}^{2/3} - 1 \right )^3}{(v_{\rm g}^2 - 1)^{3/2}} <
(2 \cos 4 \pi/9 )^{3/2} \frac{(1 - 2 \cos 4 \pi/9)^3}{(1 - 8 \cos^3 4 \pi/9)^{3/2}},
\]
where $2 \cos 4 \pi/9 = 0.3472 \cdots$.
Then,  $\partial L/\partial \xi$ is negative if
\begin{eqnarray}
&& \frac{1}{\gamma_0^2} \frac{\partial L}{\partial \xi}
< 2 v_0 (v_0^2 - 1)^3 \left [ 
- \frac{2}{(v_{\rm g}^2 -1)^{3/2}} \left \{ 
1 - \frac{1}{4} \left ( \frac{\sqrt{3}}{2} \right )^3 
\times 0.35^{3/2}( 1 - 1/3)^3 \right \} + \frac{4 \eta^2}{\sigma^2} \right ] \nonumber \\
&& < 2 v_0 (v_0^2 - 1)^3 \left [ 
- \frac{2}{(v_{\rm g}^2 -1)^{3/2}} \times 0.989993 
+ \frac{4 \eta^2}{\sigma^2} \right ] 
< 0. \nonumber
\end{eqnarray}
On the other hand, because we neglect the third order terms of $|k-k_0| \leq 1/\sigma$ 
with respect to $\omega$,
we already assumed that 
\begin{equation}
\left |  \frac{\frac{1}{3!} \frac{\partial^3 \omega}{\partial k^3} \frac{1}{\sigma^3}}
{\frac{1}{2!} \frac{\partial^2 \omega}{\partial k^2} \frac{1}{\sigma^2}} \right |
= \frac{1}{3} \left | \frac{1}{D} \frac{\partial D}{\partial k} \right | \frac{1}{\sigma}
\ll 1,
\end{equation}
where $D = \partial^2 \omega/\partial k^2 = - 4 \eta/[(2 \eta k)^2 -1]^{3/2}$.
When we use the relation $((2 \eta k)^2-1)^{-1} = v_{\rm g}^2-1$, we have
\[
\frac{4 \eta k (v_{\rm g}^2 -1) \eta}{\sigma} \ll 1
\]
Using the relation, $\frac{v_{\rm g}}{\eta k} = 2 (v_{\rm g}^2 -1)^{1/2}$, we have
\[
\frac{\eta}{\sigma} \ll \frac{1}{2 v_{\rm g} (v_{\rm g}^2-1)^{1/2}}
\]
and then we obtain
\[
\frac{4 \eta^2}{\sigma^2} \ll \frac{4}{2 v_{\rm g}^2 (v_{\rm g}^2-1)}
< \frac{2}{v_{\rm g}^3 (1- 1/v_{\rm g}^2)^{3/2}}
< \frac{2}{(v_{\rm g}^2-1)^{3/2}}.
\]
Then, we find 
\[
\frac{4 \eta^2}{\sigma^2} < \frac{2}{(v_{\rm g}^2-1)^{3/2}}
\times 0.989993.
\]
This confirms $\partial L/\partial \xi$ and then $\partial L/\partial x'$ are always negative.
This calculation shows that the slope of the transformed shape
of the original wave packet into the new coordinates with $v_0 > 1/v_{\rm g}$
is monotonically decreasing function, and its shape is drastically different
from that set in the plasma rest frame.
That is, the concept of ``wave packet" with infinite length is not 
invariant for the Lorentz
transformation, and contradicts with relativity.

\section{Lorentz invariant property of head velocity
\label{appendvheadinvariant}}

We show the frame transport property of the head velocity
of a wave train. We consider two different inertial frames
$x^\mu$ and $x^\mu{'}$, where the metric is given by
\begin{equation}
ds^2 = \eta_{\mu\nu} dx^\mu dx^\nu
= \eta_{\rho\sigma} dx^\rho{'} dx^\sigma{'}  .
\end{equation}
Then, we have the relation with respect to the transportation
coefficient between the two frames,
\begin{equation}
\eta_{\mu\nu} = \eta_{\rho\sigma} \frac{\partial x^\rho{'}}{\partial x^\mu}
\frac{\partial x^\sigma{'}}{\partial x^\nu}
= - \frac{\partial x^0{'}}{\partial x^\mu}
\frac{\partial x^0{'}}{\partial x^\nu}
+ \sum_i \frac{\partial x^i{'}}{\partial x^\mu}
\frac{\partial x^i{'}}{\partial x^\nu}  .
\label{lorentzcoeff}
\end{equation}
The transform of the 4-wavenumber $k^\mu = (\omega, \VEC{k})$
is given by
\begin{eqnarray}
\omega{'} &=& \frac{\partial x^0{'}}{\partial x^0} \omega
+ \sum_i \frac{\partial x^0{'}}{\partial x^i} k^i  , \nonumber \\
k^i{'} &=& \frac{\partial x^i{'}}{\partial x^0} \omega
+ \sum_j \frac{\partial x^i{'}}{\partial x^j} k^j  .
\end{eqnarray}
The wave head velocity in the coordinates frame $O'-x'y'z'$ is 
calculated 
\begin{eqnarray}
v_{\rm h}' &=& \lim_{\omega{'} \longrightarrow \infty} \frac{\omega{'}}{k'} 
= \lim_{\omega{'} \longrightarrow \infty}
\frac{\omega{'}}{\sqrt{\sum_i k_i'^2}}
\nonumber \\
&=& \lim_{\omega{'} \longrightarrow \infty} 
\frac{\frac{\partial x^0{'}}{\partial x^0} \omega 
+ \sum_i \frac{\partial x^0{'}}{\partial x^i} k^i}{\left [
\sum_i \left ( \frac{\partial x^i{'}}{\partial x^0} \omega 
+ \sum_j \frac{\partial x^i{'}}{\partial x^j} k^j \right )^2
\right ]^{1/2}}    \nonumber \\
&=& \frac{\frac{\partial x^0{'}}{\partial x^0} v_{\rm h}
+ \sum_i \frac{\partial x^0{'}}{\partial x^i} n^i}{\left [
\sum_i \left ( \frac{\partial x^i{'}}{\partial x^0} v_{\rm h}
+ \sum_j \frac{\partial x^i{'}}{\partial x^j} n^j \right )^2
\right ]^{1/2}}  ,
\end{eqnarray}
where we define $n^i = k^i/k$. Using Eq. (\ref{lorentzcoeff}), we derive
\[
\sum_i \left ( \frac{\partial x^i{'}}{\partial x^0} v_{\rm h}
+ \sum_j \frac{\partial x^i{'}}{\partial x^j} n^j \right )^2
= \left ( \frac{\partial x^0{'}}{\partial x^0} v_{\rm h}
+ \sum_j \frac{\partial x^0{'}}{\partial x^j} n^j \right )^2
+1 -v_{\rm h}^2  .
\]
Then, if $v_{\rm h} \leq 1$, we have $v_{\rm h}' \leq 1$.
The equality stands only if $v_{\rm h} = 1$ in a
certain frame.


\begin{table}
\caption{Microscopic variables 
($\omega_{\rm p}^{\rm e-i}$, $\omega_{\rm c}^{\rm e-i}$, $\cdots$)
and characteristic scales of phenomena of plasmas around black holes 
to examine validity of the standard resistive RMHD equations.
}
\begin{ruledtabular}
\begin{tabular}{lcccc}
    & GRB & BH X-ray binary & 
\parbox[c]{.2 \textwidth}{Supermassive BH in Galaxy} & AGN \\
\cline{1-5}
Objects    & GRB030329 & LMC X-3 & Sgr A* & M87 \\
\cline{1-5}
$M_{\rm BH}$ [$M_\odot$]\footnotemark[1]  & 3 & 10  & $2.6 \times 10^6$ & $3 \times 10^9$ \\
$\dot{M}$\footnotemark[1] & 0.1$M_\odot$ $\rm s^{-1}$ & $10^{-8} M_\odot$ $\rm yr^{-1}$ 
& $10^{-5} M_\odot$ $\rm yr^{-1}$ & $10^{-2} M_\odot$ $\rm yr^{-1}$\\
$\rho$ [$\rm g \, cm^{-3}$]\footnotemark[1] & $1.6 \times 10^{10}$ & 0.0072 & $1.5\times 10^{-6}$ & $1.3 \times 10^{-8}$ \\
$T_{\rm e}$ [K]\footnotemark[1] & $1.2 \times 10^{10}$ & $1.8 \times 10^8$ & $3.5 \times 10^5$ & $7.1 \times 10^4$ \\
$B$ [G]\footnotemark[1] & $2.7\times 10^{14}$  & $1.1\times 10^{6}$  & $1.4\times 10^2$  & 3.7 \\
\cline{1-5}
$n_{\rm e} \, [{\rm m}^{-3}]$ & $0.9 \times 10^{39}$ & $4 \times 10^{27}$ & $0.9 \times 10^{24}$ & $0.8 \times 10^{22}$ \\
$T_{\rm e}$ [eV] & $1.2 \times 10^{6}$ & $2 \times 10^{4}$ & $3.5 \times 10$ & $7 $ \\
$\nu_{\rm ei} \, [{\rm s}^{-1}]$ & $4.3 \times 10^{19}$ & $0.9 \times 10^{11}$ & $3 \times 10^{11}$ & $3 \times 10^{10}$ \\
$S_{\rm M} = \frac{\mu \omega_{\rm p}^2}{\omega \nu_{\rm ei}} $ & $5 \times 10^{14}$ & $2 \times 10^{9}$ & $3 \times 10^{8}$ & $3 \times 10^{9}$ \\
\cline{1-5}
$\omega_{\rm p}^{\rm e-i}$ [$\rm s^{-1}$] & $5.4\times 10^{21}$  & $3.6\times 10^{15}$ & $5.3\times 10^{13}$ & $4.9\times 10^{12}$\\
$\omega_{\rm c}^{\rm e-i}$ [$\rm s^{-1}$] & $2.5\times 10^{18}$  & $1.0\times 10^{10}$ & $1.3\times 10^{6}$ & $3.5\times 10^{4}$\\
\cline{1-5}
$L < L_{\rm CS}$ [m]\footnotemark[2] & 480 & 10 & $2.3 \times 10^4$ & $7.5 \times 10^5$ \\
$\tau = \omega^{-1} <  L_{\rm CS}/c$ [s]\footnotemark[2] & $1.6 \times 10^{-6}$ & $3.3 \times 10^{-8}$ & $7.7 \times 10^{-5}$& $2.5 \times 10^{-3}$ \\
\cline{1-5}
inertia of current, $\mu \frac{\omega}{\nu_{\rm ei}}$ & $8 \times 10^{-18}$ & $2 \times 10^{-7}$ & $2 \times 10^{-11}$ & $7 \times 10^{-12}$ \\
Hall effect, $\Delta \mu \frac{\omega_{\rm c}}{\nu_{\rm ei}}$ &  0.06 &  0.1 & $4 \times 10^{-6}$ & $1 \times 10^{-6}$  \\
\parbox[c]{.2 \textwidth}{thermoelectromotive force, $\leq \frac{\mu \omega_{\rm p}^2}{\nu_{\rm ei} \omega_{\rm c}}$ }
& 142 & $7 \times 10^{6}$ & $4 \times 10^{6}$ & $1 \times 10^{7}$ \\
\parbox[c]{.25 \textwidth}{momentum of charge, $\frac{2 \mu \omega_{\rm c}}{S_{\rm M} \nu_{\rm ei}}$} & $1 \times 10^{-19}$ & $6 \times 10^{-14}$ & $2 \times 10^{-17}$ & $4 \times 10^{-19}$ 
\footnotetext[1]{Data are from \citet{mckinney04t}}
\footnotetext[2]{The minimum values of $L$ and $\tau$ are estimated
by the thickness ($L_{\rm CS}$) and the light transit time 
($L_{\rm CS}/c$) of the current sheet
caused by magneto-rotational instability (MRI) in the disk.
}
\end{tabular}
\end{ruledtabular}
\end{table}

\end{document}